\documentclass[12pt]{article}
\title{Point Form Quantum Field Theory on Velocity Grids, I:  
Boson Contractions}
\author{W. H. Klink\\Department of Physics and Astronomy\\University of
Iowa, Iowa City,Iowa}
\begin{document}
\maketitle
\begin{abstract}
In constrast to discretized space-time approximations to continuum
quantum field theories, discretized velocity space
approximations to continuum quantum field theories are investigated.  A
four-momentum operator is given in terms of 
bare fermion-antifermion-boson creation and annihilation operators
with discrete indices.  In continuum quantum field theories the
fermion-antifermion creation and annihilation operators appear as
bilinears in the four-momentum operator and generate a unitary algebra. 
When the number of modes range over only a finite number
of values, the algebra is that associated with the Lie algebra of
U(2N).  By keeping N finite (but arbitrary) problems due to an
infinite Lorentz volume and to the creation of infinite numbers of bare
fermion-antifermion pairs are avoided. But even with a finite number
of modes, it is still possible to create an infinite number of bare
bosons.  We show how the full boson algebra arises as the contraction
limit of another unitary algebra that restricts the number of bare
bosons in any mode to be finite.  Generic properties of finite mode
 Hamiltonians are investigated, as are several simple models to see the
rate of convergence of the boson contraction; the possibility of fine
tuning the bare strong coupling constant is also briefly discussed.

Classification: 81R10, 81R15, 81T05, 81T25, 81T27.
\end{abstract}

\section{Introduction}

A central problem in hadronic physics is constructing nonperturbative
 solutions to QCD.  One way to obtain nonperturbative solutions is to
make space-time discrete and look for solutions on a space-time lattice. 
This is the lattice QCD program \cite{a}.

Another possibility is to discretize in momentum space;  various
groups
 have attempted to find Hamiltonian QCD solutions using instant and front
form versions of discretized quantum field theory\cite{b}.  In this
series of papers I explore quantum field theory on velocity grids, in
the context of point form quantum field theory\cite{c}.  The motivation for this
work is not only to obtain approximate nonperturbative solutions to
Hamiltonian quantum field theories and in particular Hamiltonian QCD, but
more generally to study the generic eigenvalue structure of trilinear and
boson self-coupling interactions.

In the point form \cite{d} all interactions are in the four-momentum 
operator and Lorentz transformations are kinematic.  Interactions are
introduced via vertices, products of local free fields, which are
integrated over the forward hyperboloid to give the interacting
four-momentum operator.  The natural variable that arises in point form
is the four-velocity, the four-momentum divided by the bare mass of
underlying constituents, and it is the space component of the
four-velocity that is made discrete and finite.

 The four-momentum operator
$P^{\mu}$ will be written as the sum of free and interacting four-momentum
operators,
$P^{\mu}=P^{\mu}(fr)+P^{\mu}(I)$.   To guarantee the relativistic
covariance of the theory, it is required that
\begin{eqnarray}
[P^{\mu},P^{\nu}]&=&0,\\
U_{\Lambda}P^{\mu}U_{\Lambda}^{-1}&=&(\Lambda^{-1})^{\mu}_{\nu}P^{\nu},
\end{eqnarray}
where $U_{\Lambda}$ is the unitary operator representing the Lorentz
transformation $\Lambda$.  These "point form" equations\cite{e}, in which
all of the interactions are in the four-momentum operator and the Lorentz
transformations are kinematic, lead to the eigenvalue problem
\begin{eqnarray}
P^{\mu}|\Psi_p>&=&p^{\mu}|\Psi_p>,
\end{eqnarray}
where $p^{\mu}$ is the four-momentum eigenvalue and $|\Psi_p>$ the
eigenvector of the four-momentum operator, which acts in generalized
fermion-antifermion-boson Fock spaces.   Then
the physical vacuum and physical bound and scattering states should
all arise as the appropriate solutions of the eigenvalue Eq.(3).  What is
unusual in Eq.(3) is that the momentum operator has interaction terms. 
But since the momentum and energy operators commute and can be
simultaneously diagonalized, they have common eigenvectors.  One of the
important properties of the point form is that the Lorentz generators
have no interactions, so that global Lorentz transformations on operators
and states are simple and explicit.

Excluding boson self-interactions, all of the
fundamental particle interactions have the form of bilinears
in fermion and antifermion creation and annihilation operators times
terms linear in boson creation and annihilation operators.  For example
QED is a theory bilinear in electron and positron creation and
annihilation operators and linear in photon creation and annihilation
operators.  The well-known nucleon-antinucleon-meson interactions are of
this form as are the weak interactions.  These interactions differ of
course in the way the fermions are coupled to the bosons, including the
way in which internal symmetries are incorporated.  For QCD,
because of the
$SU(3)_{color}$ symmetry which generates gluon self-coupling terms, the
gluon sector is no longer linear in creation and annihilation operators; 
this is also the case for the weak interactions and with
gravitons; since gravitons carry energy and momentum, they also can couple
to themselves.  But even with bosonic self-interactions, the coupling of
bosons to fermions is trilinear.

If $a^{\dagger}, b^{\dagger}$ and $c^{\dagger}$ denote
respectively, fermion, antifermion and boson creation operators, the
aformentioned trilinear interactions can all be written as
$(a^{\dagger}+b)(a+b^{\dagger})(c^{\dagger}+c)$, while the "relativistic
energy" terms are of the form $a^{\dagger}a-bb^{\dagger}+c^{\dagger}c.$ 
Written in this way the fermion-antifermion bilinears $a^{\dagger}a,
bb^{\dagger},a^{\dagger}b^{\dagger}$, and $ba$ close to form a Lie
algebra which is related to the Lie algebra of the unitary groups. 
Similarly the boson operators $c^{\dagger}c, c^{\dagger}$ and $c$ close
to form a Lie algebra related to the semidirect product of unitary groups
with the Heisenberg group.  Then the aforementioned interactions can all
be viewed as arising from an algebra of operators generated from these two Lie algebras.

One of the main problems that arises in solving continuum
field theory eigenvalue equations such as Eq.(3) is that the interacting
four-momentum operator takes elements out of the generalized
fermion-antifermion-boson Fock space.  Difficulties arise in three ways,
from the infinite Lorentz volume, from the possibility of creating
infinite numbers of bare fermion-antifermion pairs, and from the
possibility of creating infinite numbers of bare bosons.  If the number
of fermion-antifermion velocity modes is made finite, the first two kinds
of problems can be avoided.  As will be shown, for a finite number of
modes, N,  there is an underlying fermion-antifermion symmetry generated
by the Lie algebra of the group U(2N). With such a group structure it is
possible to define an inductive limit as N goes to infinity \cite{dd};  
inductive limits will be studied in a following paper. Appendix B shows
that for finite approximations to trilinear interactions, the ground state
energy varies linearly with the bare coupling constant (for large values
of the coupling constant), with the slope being negative.

 But even with a finite number N of modes, it is still
possible to have indefinitely large numbers of bare bosons in each mode. 
One of the main goals of this paper is to show that the algebra of bosonic
operators, generated by the semidirect product of unitary with Heisenberg algebras,
can be given as a contraction limit of another unitary algebra; 
before the contraction limit is taken, the number of bare bosons in any
mode is finite.  Then the eigenvalue problem, Eq.(3) becomes a problem in
diagonalizing matrices.  The maximum number
of bare bosons is controlled by a number M, which, going to infinity as
the contraction parameter goes to zero, gives the full boson algebra.

For theories like Hamiltonian QCD there are also boson self-coupling
terms.  In that case the $c^\dagger c$ terms in the free four-momentum
operator are supplemented by self-energy terms.  But the boson
contraction limit is still valid with these self-coupling terms, only
now the matrices to be diagonalized are more complicated. A simple one
mode example is given in section 5 to see how fast the boson
contraction converges; boson self-coupling terms are then added and the
possibility of fine tuning the bare strong coupling constant is discussed.
This section also analyzes a simple few mode isospin model.  To further
examine the nature of bosonic contraction, Appendix A discusses a
simple exactly solvable bosonic model.

Section 2 provides an introduction to point form relativity; starting
with trilinear interactions, the existence of a unitary symmetry arising
from fermion bilinears is discussed in section 3.  Section 4 deals with
bosons and boson contractions.

\section{Point Form Vertex Interactions}  

The starting point in this paper for approximating the quantum field
theory eigenvalue equation are the massive one-particle representations\cite{f} of
the Poincar\'{e} group, with four-velocity states
 $|v,\sigma>$, where
$v\cdot v=1$, and the four-momentum is $p=mv$; $\sigma$ is the 
spin projection ranging between $-j\le\sigma\le j$.  $m$ and $j$ are the
mass and spin respectively.  (For massless particles, $v\cdot v=0$ and is
treated separately)

Under a Poincar\'{e}
transformation,
\begin{eqnarray*}
U_a|v,\sigma>&=&e^{ip\cdot a} |v,\sigma>\\
U_{\Lambda}|v,\sigma>&=&\sum |\Lambda v,\sigma^{'}>D^j_{\sigma^{'},\sigma}
(R_W(v,\Lambda)),
\end{eqnarray*}
where $a$ is a four-translation, $D^j_{\sigma^{'},\sigma}()$ is an $SU(2)$
matrix element for spin
$j$ and $R_W(v,\Lambda)$ is a Wigner rotation, an element of the rotation
group $SO(3)$ defined by
\begin{eqnarray*}
R_W(v,\Lambda):&=&B^{-1}(\Lambda v)\Lambda B(v).
\end{eqnarray*}
$B(v)$ is a boost, a Lorentz transformation satisfying $v=B(v)v^{rest}$,
with $v^{rest}=(1,0,0,0)$.  Boosts are always  
canonical spin boosts \cite{bb}.

From such one-particle representations, a many-body theory can be
formulated by introducing creation and annihilation operators with the
same transformation
properties as one-particle states;  these generate multiparticle 
states from the Fock vacuum:
\begin{eqnarray}
|v,\sigma>&=&a^{\dagger}(v,\sigma)|0>,\\
{[}a(v,\sigma),a^{\dagger}(v^{'},\sigma^{'})]_{\pm}&=&2v_0\delta^3(v-v^{'})
\delta_{\sigma,\sigma^{'}}\\
U_aa(v,\sigma)U_a^{-1}&=&e^{-ip.a}a(v,\sigma)\\
U_\Lambda a(v,\sigma)U_\Lambda^{-1}&=&\sum a(\Lambda v,\sigma^{'})
D^j_{\sigma^{'},\sigma}(R_W(v,\Lambda))^{\ast},
\end{eqnarray}
where the $\pm$ denotes commutator or anticommutator for bosons or
fermions respectively. By using velocity rather than momentum
variables, all operators are dimensionless.

Let $a_i,b_i,c_k$ denote respectively bare fermion, antifermion, and
boson annihilation operators where the indices stand for both space-time
and internal
 variables such as charge or isospin. 
Then the free four-momentum operator can be written as
\begin{eqnarray}
P^{\mu}(fr):&=&m\sum \int dv
v^{\mu}(a^{\dagger}_ia_i+b^{\dagger}_i b_i+\kappa c^{\dagger}_kc_k),
\end{eqnarray}
where $dv:=\frac{d^3v}{v_0}$ is the Lorentz invariant measure, $\kappa$
is a dimensionless relative bare boson mass parameter and $m$ is a
constant with the dimensions of mass; its value is determined by relating
a physical mass such as the nucleon mass to the dimensionless eigenvalue
of the corresponding stable particle.  The indices $i$ and $k$ contain both the continuous velocity modes and finite spin projection and internal symmetry degrees of freedom.   Because of the transformation
properties of the creation and annihilation operators inherited from the
one particle states, the free four-momentum operator, as defined in
Eq.(8), satisfies the point form equations (1) and (2)\cite{c}.

Interacting
four-momentum operators  are generated from vertices, products of free
field operators themselves made from creation and annihilation operators. 
Denote a vertex operator by $V(x)$, where $x$ is a space-time
point. 
Then the interacting four-momentum operator is obtained by
integrating the vertex operator over the forward hyperboloid:
\begin{eqnarray}
P^{\mu}(I):&=&g\int d^4 x\delta(x\cdot x-\tau^2)\theta(x_0) x^{\mu} V(x)\\
&=&g\int dxx^{\mu}V(x),\nonumber
\end{eqnarray}
with $g$ a coupling constant. 

The vertex
operator is required to be a scalar density under Poincar\'{e}
transformations and have locality properties.  That is,
\begin{eqnarray}
U_a V(x) U_a^{-1}&=&V(x+a)\\
U_\Lambda V(x)U_{\Lambda}^{-1}&=&V(\Lambda x);\\
{[}V(x),V(y)]&=&0
\end{eqnarray}
if $(x-y)^2$ is spacelike.  Here $U_a=e^{-iP(fr)\cdot a}$ is the free
four-translation operator. 

Making use of the fact that if $x$ and $y$ are two time-like four-vectors
with the same length, so that their difference is space-like, it follows
that $[P^{\mu},P^{\nu}]=0$.  Also, from the Lorentz transformation
properties of the vertex given in Eq.(11), it follows that the
interacting four-momentum operator transforms as a four-vector.  Thus the
interacting four-momentum operator also satisfies the point form
equations (1) and (2).

If the free four-translations are made infinitesimal, then
$[P^{\nu}(fr),P^{\mu}(I) ]=\int dx x^{\mu}\frac{\partial}{\partial
x_{\nu}}V(x)$, so that
\begin{eqnarray*}
{[}P^{\nu}(fr),P^{\mu}(I)]-[P^{\mu}(fr),P^{\nu}(I)]&=&g\int dx(x^{\mu}
\frac{\partial}{\partial x_{\nu}}-x^{\nu}\frac{\partial}{\partial
 x_{\mu}})V(x)\nonumber\\
&=&0;\\
{[}P^\mu(fr)+P^\mu(I),P^\nu(fr)+P^\nu(I)]&=&[P^\mu,P^\nu]\\
&=&0,
\end{eqnarray*}
which means the total four-momentum operator, the sum of free and
interacting four-momentum operators, also satisfies the point form
equations.

As stated in the introduction, 
the trilinear vertex is assumed to be bilinear in fermion-antifermion
creation and annihilation operators, and linear in boson creation and
annihilation operators.  That is, such vertices have the general
form
$V\sim(a^{\dagger}+b)(a+b^{\dagger})(c+c^{\dagger})=(a^{\dagger}a
+bb^{\dagger}+a^{\dagger}b^{\dagger}+ba)(c+c^{\dagger})$
so the interacting four-momentum operator has the general form

\begin{eqnarray*}
P^{\mu}(I)&=&g\sum\int
dv_1dv_2dv(X^{\mu}_{11}(k)_{i_1i_2}a^{\dagger}_{i_1}a_{i_2}c_k+
X^{\mu}_{22}(k)_{i_1i_2}b_{i_1}b^{\dagger}_{i_2}c_k\nonumber\\
&&+X^{\mu}_{12}(k)_{i_1i_2}a^{\dagger}_{i_1}b^{\dagger}_{i_2}c_k+
X^{\mu}_{21}(k)_{i_1i_2}b_{i_1}a_{i_2}c_k+hc);
\end{eqnarray*}
here $X^{\mu}_{11}(k)_{i_1i_2}=F^{\mu}(v_1-v_2-\kappa v)
M_{11}(i_1i_2k)$ and \\$F^{\mu}(u):=\int d^4x\delta(x\cdot
x-\tau^2)x^\mu e^{ix\cdot u}$ comes from locality.
$M_{11}(i_1i_2k)$ are spinor and internal symmetry matrices coming
from the free fields, dependent on how the fermions and antifermions are
coupled to the bosons.
The other three $X$'s have a similar form.
 
Now all quantities except the coupling constant are dimensionless, so g
has the dimensions of mass.  Write $g=m\alpha$ and divide by $m$. 
Then all four-momentum operators are dimensionless.

Further, all fermion-antifermion terms are bilinears with the common
factor
$c_k$ or
$c^{\dagger}_k$.  So write
\begin{eqnarray}
P^{\mu}(I)&=&\alpha\sum\int dv(\mathcal{A}(X^{\mu}_k)
c_k+\mathcal{A}(X^{\mu}_k)^{\dagger}c^{\dagger}_k),
\end{eqnarray}
where
$\mathcal{A}(X^\mu_k):=(a^{\dagger}_{i_1},b_{i_1})(X^{\mu}_k)_{i_1i_2}
(a_{i_2},b^{\dagger}_{i_2})^T$.

In these variables,the eigenvalue problem, Eq.(3), to be solved for trilinear interactions is
\begin{eqnarray}
(P^{\mu}_{F}(fr)+\sum\int dv(\kappa v^{\mu}c^{\dagger}_k
c_k&&\nonumber\\+\alpha
\mathcal{A}(X^{\mu}_k)c_k+\alpha
\mathcal{A}(X^{\mu}_k)^{\dagger}c^{\dagger}_k))|\Psi_{\lambda}>
&=&\lambda^{\mu}|\Psi_{\lambda}>
\end{eqnarray}

This eigenvalue equation is covariant, so Lorentz covariance can be used
to write
$\lambda^{\mu}=(\lambda,0,0,0)$.  Then  the zeroth component of
the eigenvector equation is
\begin{eqnarray}
P^{0}_{F}(fr)+\sum\int dv(\kappa v^{0}c^{\dagger}_k
c_k\nonumber\\
+\alpha\mathcal{A}(
X^{0}_k)c_k+\alpha\mathcal{A}(X^{0}_k)^{\dagger}c^{\dagger}_k))
|\Psi_{\lambda}>
&=&\lambda|\Psi_{\lambda}>
\end{eqnarray}

In particular the vacuum eigenvalue problem is characterized by
$P^0|\Omega>=0$ and $U_{\Lambda}|\Omega>=|\Omega>$.  If the point form
Eq.(2) holds, then
\begin{eqnarray*}
U_{\Lambda}P^0|\Omega>&=&U_{\Lambda}P^0U_{\Lambda}^{-1}
U_{\Lambda}|\Omega>\\
&=&((\Lambda^0_0)^{-1}P^0+(\Lambda^0_i)^{-1}P^i)|\Omega>\\
&=&(\Lambda^0_i)^{-1}P^i|\Omega>\\&=&0,
\end{eqnarray*}
which implies that the momentum operator acting on the physical vacuum
also gives zero, as required.  Thus it suffices to analyze the eigenvalue
problem, Eq.(3) only for the zero component, $\mu=0$.

A fundamental difficulty in solving the eigenvalue problem, Eq.(15), is
that the interaction term takes elements out of
the Fock space.  To remedy this problem, orthonormal bases in the one
particle spaces are truncated so there are N basic modes, which include the finite
 velocity modes, as well as spin projection and internal symmetry
modes, and as N gets larger, the continuum (inductive\cite{dd}) limit is
approached.

To distinguish between the continuum energy operator in Eq.(15) and its
finite approximation, the fundamental operator for trilinear interactions will
be (a Hamiltonian) denoted by $H$, made out of creation and annihilation
operators with a finite number of modes,  whose form mimics Eq.(15):
\begin{eqnarray}
H&=&\sum e_i(a^{\dagger}_ia_i+b^{\dagger}_ib_i+\kappa c^{\dagger}_kc_k)
+\alpha\sum\mathcal{A}(X_k)c_k+\mathcal{A}(X_k^{\dagger})
c^{\dagger}_k\\
&=&\sum e_i+\sum e_i(a^{\dagger}_ia_i-b_ib^{\dagger}_i+\kappa
c^{\dagger}_ic_i)+\alpha(\mathcal{A}(X_i)c_i+\mathcal{A}(X^{\dagger}_i)
c^{\dagger}_i)\\
&=&\sum e_i+\mathcal{A}(E)+\kappa\sum e_i
c^{\dagger}_ic_i+\alpha\sum(\mathcal{A}(X_i)c_i
+\mathcal{A}(X^{\dagger}_i) c^{\dagger}_i),\\
E:&=&diag(e_1, e_2,...,e_N, -e_1,-e_2,...,-e_N),
\end{eqnarray}
where the discrete "energy" $e_i=\sqrt{1+v_i^2}$, and the $\mathcal{A}()$
notation emphasizes the unitary structure developed in the next section.

\section{ U(2N) Fermionic Structure}

The fermion-antifermion creation and annihilation operators appear
in the Hamiltonian, Eq.(18), only in the combinations
$a^{\dagger}_ia_j,b_ib^{\dagger}_j,a^{\dagger}_ib^{\dagger}_j$
and $b_ia_j$.  Indices range from 1 to N (number of fermion modes).
Thus, consider the correspondence
$a^{\dagger}_i\rightarrow A^{\dagger}_i, a_i\to A_i,
b_i\to A^{\dagger}_{i+N}, b^{\dagger}_i\to A_{i+N}$, where the indices on
the A's range between 1 and 2N.
Then the bilinears $A^{\dagger}A$ generate the
Lie algebra of $U(2N)$, with commutation relations 
\begin{eqnarray}
[A^{\dagger}_{\alpha}A_{\beta},
A^{\dagger}_{\mu}A_{\nu}]=A^{\dagger}_{\alpha}A_{\nu}\delta_{\beta,\mu}
-A^{\dagger}_{\mu}A_{\beta}\delta_{\alpha,\nu}.
\end{eqnarray} 
 For example
\begin{eqnarray*}
{[}b_ia_j,
a^{\dagger}_kb^{\dagger}_l]&=&[A^{\dagger}_{i+N}A_j,A^{\dagger}_k
A_{l+N}]\nonumber\\&=&
A^{\dagger}_{i+N}A_{l+N}\delta_{j,k}-A^{\dagger}_kA_j\delta_{i,l}\nonumber\\
&=&b_ib^{\dagger}_l\delta_{j,k}-a^{\dagger}_ka_j\delta_{i,l},
\end{eqnarray*}
as required.

Again use the definition
$\mathcal{A}(X):=A^{\dagger}_{\alpha}X_{\alpha\beta}A_\beta,1 \le
\alpha,\beta\le 2N$; then all fermionic terms in $H$ are of this
form;  for example, the free fermionic energy is
\begin{eqnarray*}
H_F(fr)&=& \sum e_i+\sum e_i(a^{\dagger}_i a_i-b_ib^{\dagger}_i)\\
&=&\sum e_i+\sum\mathcal{A}(E).
\end{eqnarray*}

It should be noted that $bb^{\dagger}$ and not $b^{\dagger}b$ is the
proper element in the Lie algebra of $U(2N)$, Eq.(20).  Further, there is no
longer any trace of fermion and antifermion anticommutation relations; 
all of the fermionic structure is given by the Lie algebra of $U(2N)$. 
Only the representations of this Lie algebra can be traced back to the
underlying fermionic structure.

The representations of $U(2N)$, in Gelfand-Zetlin labeling \cite{g},
 are all the antisymmetric representations, written
(1,...,1,0,...0), of length 2N, with the identity representations given
by all zeroes or all ones.  The antisymmetric Fock space is the direct
sum of all these irreducible representation spaces, and is of dimension
$2^{2N}$.

But a more convenient way of labelling the fermionic
representation spaces is with the "baryon number" operator.  Define
$\mathcal{B}+N:=\sum
(a^{\dagger}_ia_i+b_ib^{\dagger}_i)=\sum\mathcal{A}(I)$, a first order
Casimir operator with eigenvalues from -N to
+N.  Each integer corresponds to a given irreducible representation. 
For example the $\mathcal{B}=0$ sector corresponds to the Gelfand-Zetlin
label with equal numbers of zeroes and ones. 

In the $\mathcal{B}=0$ sector, fermionic states can be written as
$|\mu_1\mu_2...\mu_N>$, where the $\mu$'s form a shuffle ranging between 1 and 2N\cite{h};  a concrete realization is given by minors of
determinants, 
$\Delta^{1.....N}_{\mu_1..\mu_N}(z)$, with $z$ a 2N$\times$ 2N matrix\cite{g}.
Writing states as $|\mu_1\mu_2...\mu_N>$ is equivalent to writing 
$a^{\dagger}_{i_1}b^{\dagger}_{j_1}...a^{\dagger}_{i_n}b^{\dagger}_{j_n}
|0>$; in this notation the Fock vacuum is $|0>=|N+1...2N>\sim
\Delta^{1........N}_{N+1..2N}(z)$, while the highest state is 
$a^{\dagger}_{i_1}b^{\dagger}_{j_1}...a^{\dagger}_{i_N}b^{\dagger}_{j_N}
|0>=|1...N>\sim
\Delta^{1...N}_{1...N}(z)$. 

 Since
fermion-antifermion operators appear only as bilinears in the
Hamiltonian, Eq.(18), a fundamental operator identity which makes such a
basis representation useful is
\begin{eqnarray}
A^{\dagger}_{\alpha}A_{\beta}|\mu_1...\mu_N>&=&\delta_{\mu_1
\beta}|\alpha...\mu_N>+...\delta_{\mu_N\beta}|\mu_1...\alpha>
\end{eqnarray}

All other baryon number subspaces can be  written as
$|\mu_1\mu_2...\mu_k>, k=1...2N$, and an operator identity similar to
Eq.(21) applies also to these subspaces.
  
To get from the $\mathcal{B}$=0 subspace to other baryon number
subspaces, it is possible to use products of operators,
$A^{\dagger}_{\mu_1}...A^{\dagger}_{\mu_k}$.  Though single operators
anticommute among themselves, they satisfy commutation relations with the
bilinears:
\begin{eqnarray}
{[}A^{\dagger}_\alpha A_\beta,A^{\dagger}_\mu]
&=&\delta_{\beta\mu}A^{\dagger}_\alpha\\
{[}\mathcal{A}(Y),A_\mu^{\dagger}]&=&Y_{\alpha\mu}A^{\dagger}_\alpha\\
{[}A^{\dagger}_\alpha A_\beta,A_\mu]
&=&-\delta_{\alpha\mu}A_\beta.
\end{eqnarray}

Here the contrast can be seen with the representation structure of the
usual fermion anticommutation relations.  The fermionic irreducible
representation spaces of U(2N) are given by the baryon number operator
eigenvalues, ranging between -N and +N.  Since the Hamiltonian is
bilinear in fermion-antifermion operators, it never mixes these spaces. 
However, single fermion operators do mix these spaces, resulting in
there being only one irreducible representation for the fermion
commutation relations, which is the direct sum of all the baryon number
subspaces.  In the limit when N goes to infinity, the representation
structure radically changes;  it is this inductive limit\cite{dd} that will be
studied in the next paper.

\section{Bosonic Contractions}
Even when the number of modes N is finite, it is still possible to
have an infinite number of bosons in each mode.  In this section
we show how to contract a unitary Lie algebra \cite{i} to the full boson
algebra;  since the irreducible representations of a unitary
algebra are finite dimensional, the contraction is done in such a
way that as the contraction parameter goes to zero, the
irreducible representation goes to infinity.

Consider then the bosonic Lie
algebra consisting of creation and annihilation operators,
$c_i^{\dagger}, c_i$. Adjoin to this the elements
$L_{ij}:=c_i^{\dagger}c_j$, so that the commutation relations of the four
elements are
\begin{eqnarray}
{[}L_{ij},c^{\dagger}_k]&=&c^{\dagger}_i\delta_{jk}\\
{[}L_{ij}, c_k]&=&-c_i\delta_{jk}\\{[}L_{ij},I]&=&0\\
{[}c_i,c_j^{\dagger}]&=&I\delta_{ij}\\
{[}c_i,I]&=&0
\end{eqnarray}
The commutation relations are those of the semidirect product of the
unitary algebra with the Heisenberg algebra. 

For each mode this gives
\begin{eqnarray}
[L,c^{\dagger}]&=&c^{\dagger}\\
{[}L,c]&=&-c\\
{[}c,c^{\dagger}]&=&I,
\end{eqnarray}
which is the usual harmonic oscillator algebra for each mode.

Consider next a U(2) Lie algebra, with elements $J_1,J_2,$ and
$J_{\pm}$ and the following commutation relations:
\begin{eqnarray}
{[}J_1,J_2]&=&0\\
{[}J_1,J_{\pm}]&=&\pm J_{\pm}\\
{[}J_2,J_{\pm}]&=&(-)\pm J_{\pm}\\
{[}J_-,J_+]&=&J_2-J_1
\end{eqnarray}

Now modify the basis of this Lie algebra by defining
$\tilde{J}_{\pm}:=\rho J_{\pm}$ and $\tilde{J}_2:=\rho^2 J_2$, with $\rho$
a positive number;  then the Lie algebra becomes
\begin{eqnarray}
{[}J_1,\tilde{J}_2]&=&0\\
{[}J_1,\tilde{J}_{\pm}]&=&\pm \tilde{J}_{\pm}\\
{[}\tilde{J}_2,\tilde{J}_{\pm}]&=&\pm (-\rho^2)\tilde{J}_{\pm}\\
{[}\tilde{J}_-,\tilde{J}_+]&=&\tilde{J}_2-\rho^2 J_1\\
\end{eqnarray}
In the contraction limit when $\rho\rightarrow0$ this Lie algebra agrees
with one mode bosonic algebra ($\tilde{J}_2$ plays the role of the identity
operator)

Next consider a concrete realization of the U(2) Lie algebra, with
\begin{eqnarray}
J_1&\rightarrow&z\frac{\partial}{\partial z}\\
J_2&\rightarrow&w\frac{\partial}{\partial w}\\
J_+&\rightarrow&z\frac{\partial}{\partial w}\\
J_-&\rightarrow&w\frac{\partial}{\partial z}
\end{eqnarray}
on the holomorphic Hilbert space of two complex variables\cite{g}.  Then the
bosonic representations, labelled (M,0), are the homogeneous
polynomials of degree M, with an orthonormal basis given by
$<z,w|M,n>=\frac{z^{n}w^{M- n}}{\sqrt{n!(M-n)!}}$.  In order that such a
basis be holomorphic, the total number of bosons in any mode is
restricted by M.  Further
\begin{eqnarray}
<z,w|\tilde{J}_+|M,n>&=&\frac{\rho
(M-n)z^{(n+1)}w^{(M-n-1)}}{\sqrt{n!(M-n)!}}\nonumber\\
&=&\sqrt{n+1}\sqrt{\rho^2(M-n)}<z,w|M,n+1>.
\end{eqnarray}
When $M\rightarrow\infty,\rho\rightarrow 0$ such that $M\rho^2=1$,
the usual boson calculus result is recovered.

This can all be generalized to N modes;  now the Lie algebra is U(N+1)
with generators 
\begin{eqnarray}
J_{ij}&\rightarrow& z_i\frac{\partial}{\partial z_j}\\
J_i^+&\rightarrow&z_i\frac{\partial}{\partial w}\\
J_i^-&\rightarrow& w\frac{\partial}{\partial z_i}\\
J_2&\rightarrow&w\frac{\partial}{\partial w}
\end{eqnarray}
As with one mode, $\rho J_i^{-}\rightarrow c_i, 
\rho^2J_2\rightarrow I$, as $\rho\rightarrow0$.

Bosonic representations of U(N+1) are written as (M,0,...0)\cite{g}; polynomial
realizations of basis states are given by
\begin{eqnarray}
<z,w|M\vec{n}>&=&\frac{z_1^{n_1}...z_N^{n_N} w^{(M-\sum
n_i)}}{\sqrt{n_1!...n_N! (M-\sum n_i)!};}\\
J_{ii}|M\vec{n}>&=&n_i|M\vec{n}>\\
J_2|M\vec{n}>&=&(M-\sum n_i)|M\vec{n}>\\
J_i^-|M\vec{0}>&=&0\\
J_i^+|M,max>&=&0.
\end{eqnarray}

The idea now is to replace a Hamiltonian given in terms of boson creation
and annihilation operators with the $\tilde{J}$ operators.  For example,
the Hamiltonian, eq.(17), is replaced by
\begin{eqnarray}
H_M&=&\sum e_i+\sum e_i(a^{\dagger}_ia_i-b_ib^{\dagger}_i+\kappa
\tilde{J}_{ii})+\alpha(\mathcal{A}(X_i)\tilde{J}^-_i+\mathcal{A}(X^{\dagger}_i)
\tilde{J}^+_i),
\end{eqnarray}
and in the contraction limit the eigenvalues of $H_M$ should pass over to
the eigenvalues of $H$.  An example of an exactly solvable model without fermions is given
in Appendix A.

\section{Modeling the QCD Vacuum: A Simple One Mode Model}

Up to this point only trilinear interactions have been discussed in this paper.  But, as shown in
Appendix B, the ground state for Hamiltonians of the form given in Eq.(18) as a
function of the bare coupling constant $\alpha$ goes as -$|constant|\alpha$;  
as discussed in the conclusion, it
is not possible to add an arbitrary constant to the Hamiltonian, so there is no zero energy
ground state for nonzero $\alpha$.  This means it is necessary to add other terms
to the model Hamiltonian.  One of the goals of this series of papers is to examine
inductive limits of the QCD Hamiltonian.  Thus it is natural to add boson
self-coupling terms to the Hamiltonian with trilinear interactions.

But before examining a model Hamiltonian with both trilinear and boson
self-coupling terms, it is necessary to investigate the convergence rate
of boson contractions.
The previous section showed that boson contraction converges to the
full infinite dimensional limit.  What is not clear is how fast boson contraction
converges.  A way to study this question is to examine the quantum anharmonic
oscillator as a one mode field theory.

 Thus, consider the Hamiltonian 
\begin{eqnarray}
H&=&\frac{1}{2}(x^2+p^2)+\alpha^2 x^4\nonumber\\
&=&c^{\dagger} c+\alpha^2(c+c^{\dagger})^4 
\end{eqnarray}

The finite number of bosons Hamiltonian is given by generators of the 
U(N+1)=U(2) Lie algebra:
\begin{eqnarray}
H_M&=&J_1+\alpha^2(\tilde{J}_++\tilde{J}_-)^4\\
&=&\frac{M-J_z}{2}+\alpha^2\rho^4 J_x^4\\
&=&\frac{M+J_x}{2}+\frac{\alpha^2}{M^2}J_z^4,
\end{eqnarray}
and in the contraction limit, the eigenvalues of $H_M$, Eq.(58) should
converge to the eigenvalues of $H$, Eq.(57).  In Eq.(59) the Lie algebra
basis of U(2) has been written in a U(1)xSU(2) Lie algebra basis, and in Eq.(60), the
contraction parameter has been eliminated by writing $\rho=\frac{1}{M^2}$.
The goal is to numerically find the lowest eigenvalues for fixed coupling
as M gets large. Using an  SU(2) Lie algebra automorphism to interchange
the x and z generators generates a tridiagonal matrix in the basis gives
in Eq.(46).  Reference\cite{k} shows that the true eigenvalue is approached for
M about 100.

  Next consider a one mode problem coupling fermions and bosons, with
no quartic boson selfcoupling. The Hamiltonians are now
\begin{eqnarray}
H&=&1+(a^{\dagger}a-bb^{\dagger})+c^{\dagger}c+\alpha(\mathcal{A}(X)c+
\mathcal{A}(X^{\dagger})
c^{\dagger})\\
&=&1+\mathcal{A}(E)+c^{\dagger}c+\alpha(\mathcal{A}(X)c+
\mathcal{A}(X^{\dagger})c^{\dagger};\\
H_M&=&1+\mathcal{A}(E)+J_1+\alpha(\mathcal{A}(X)\tilde{J}_-+
\mathcal{A}(X^{\dagger})\tilde{J}_+)\\
E&=&\left[\begin{array}{cc}1&0\\0&-1\\
\end{array}\right],  X=\left[\begin{array}{cc}1&1\\1&1\\
\end{array}\right]. 
\end{eqnarray}

For $\mathcal{B}=0$ the fermion space is two dimensional $(|0>=|2>,
a^{\dagger}b^{\dagger}|0>=|1>$ in the notation of section 3) and the boson
space is M+1 dimensional. When $H_M$ is diagonalized, the lowest
eigenvalue for small values of M have been calculated\cite{l}. 
The behavior of the lowest eigenvalue for any of the M values
is linear decreasing with respect to the bare coupling constant, 
confirming the general behavior for a trilinear coupling that the
ground state always decreases linearly in $\alpha$ for sufficiently 
large $\alpha$.

Finally, if the boson self-coupling term, the anharmonic term in Eq.(58) is
added to the Hamiltonian, Eq.(63), we have a one mode model of trilinear
coupling with a quartic boson self-coupling, a simple QCD one mode model:
\begin{eqnarray}
H_M^{QCD}&=&1+\mathcal{A}(E)+J_1+\alpha(\mathcal{A}(X)\tilde{J}_-+
\mathcal{A}(X^{\dagger})\tilde{J}_+)\nonumber\\
&&+\alpha^2(\tilde{J}_-+\tilde{J}_+)^4;
\end{eqnarray}
the lowest eigenvalue for several small values of M, as a function of the
bare coupling parameter have also been computed \cite{l}.  For odd values
of M, the ground state eigenvalue as a function of the bare coupling
parameter passes through zero (when M is even there are spurious zeroes)
;  this raises the possibility of fine
tuning the bare coupling parameter, as discussed in the conclusion.

\section{ Fermion-Boson Isospin Model}

In this section we analyze the ground state structure of a few mode
system, generated by isospin internal symmetry and only one velocity
mode. The system will consist of isospin 1 "pions" coupled to isospin
1/2 "nucleons" and "antinucleons".  The Hamiltonian will have "kinetic
energy", trilinear coupling, and finally, "pion self-coupling" terms.

To begin, the bosonic isospin operators are generated from a U(N+1)=U(4)
bosonic symmetry, enumerated as 1,0,-1:
\begin{eqnarray}
I_3^B&=&J_{11}-J_{-1,-1} (=z_1\frac{\partial}{\partial
z_1}-z_{-1}\frac{\partial}{\partial z_{-1}})\\
I_+^B&=&\sqrt{2}(J_{10}+J_{0,-1})(=\sqrt{2}(z_1\frac{\partial}{\partial
z_0}+z_0\frac{\partial}{\partial z_{-1}}), 
\end{eqnarray}
which then satisfy the isospin algebra,
${[}I_3^B,I^B_+]=I_+^B,$ and $[I_+^B,I_-^B]=2I_3^B$.

The bosonic states in the U(4) algebra in
a holomorphic basis are given as $|M,n_+,n_0,n_->=\frac{z_1^{n_+}
z_0^{n_0} z_{-1}^{n_{-}} w^{M-\sum n_k}}{\sqrt{n_+! n_0!n_-!(M-\sum
n_k)!}}$; the no-pion state is given by M=0.

The M=1 representation is four-dimensional, with one no-pion state
(isospin 0) and three one-pion states (isospin 1):
  
  $|M=1, 000>=w$,
isospin 0

$|M=1, 100>=z_1$, isospin 1, (+1),

$|M=1, 010>=z_0$, isospin 1, (0),

$|M=1, 001>=z_{-1}$, isospin 1, (-1).

The M=2 representation is 10 dimensional, with isospin 2 and 1 states, and
two isospin 0 states;  only the 0 and 1 states are listed here:

$|M=2,I=0>=\frac{w^2}{\sqrt{2}}$,

$|M=2,I=0>=\frac{1}{\sqrt{2}}z_1z_{-1}-
\frac{1}{2\sqrt{2}}z_0^2$,

$|M=2, I=1>= wz_1,wz_0,wz_{-1}$\\

More generally the isospin 0 states can be written as\\  $|I=0>=\sum_m
\alpha_m|M=2n, n-m, 2m, n-m>$

In a similar fashion the proton and neutron give 2 modes, which, along
with their antiparticles, give the unitary group U(2N)=U(4).  In terms of
creation and annihilation operators the four particle states can be
written as
\begin{eqnarray*}
|p>&=&a_1^{\dagger}|0>, |\bar{p}>=b_1^{\dagger}|0>,\\
|n>&=&a_2^{\dagger}|0>, |\bar{n}>=b_2^{\dagger}|0>.
\end{eqnarray*}
Then the fermion isospin structure has the form:
\begin{eqnarray}
I_3^F&=&\frac{1}{2}(a_1^{\dagger} a_1-a_2^{\dagger} a_2+b_1b_1^{\dagger}
-b_2b_2^{\dagger})\nonumber\\
&=& \frac{1}{2}(A_1^{\dagger} A_1-A_2^{\dagger}
A_2+A_3^{\dagger}A_3-A_4^{\dagger}A_4)\\
&=&\mathcal{A}\left[\begin{array}{cc}\tau_3&0\\0&\tau_3\\
\end{array}\right], \tau_3= \frac{1}{2}\left[\begin{array}{cc}1&0\\0&-1\\
\end{array}\right]\\
I_+^F&=&A_1^{\dagger}A_2+A_3^{\dagger}A_4=\mathcal{A}
\left[\begin{array}{cc}\tau_+&0\\0&\tau_+\\
\end{array}\right],\tau_+=\left[\begin{array}{cc}0&1\\0&0\\
\end{array}\right]\\
{[}I_3^F,I_+^F]&=&[\mathcal{A}\left[\begin{array}{cc}\tau_3&0\\0&\tau_3\\
\end{array}\right],\mathcal{A}\left[\begin{array}{cc}\tau_+&0\\0&\tau_+\\
\end{array}\right]]=I_+^F\\
{[}I_+^F,I_-^F]&=&[\mathcal{A}\left[\begin{array}{cc}\tau_+&0\\0&\tau_+\\
\end{array}\right],\mathcal{A}\left[\begin{array}{cc}\tau_-&0\\0&\tau_-\\
\end{array}\right]]=2I_3^F
\end{eqnarray}
and the action of the operators on the primitive states is given by\\
 $I_3^F|p>=\frac{1}{2}|p>, I_3^F|n>=-\frac{1}{2}|n>$ for B=1 and by\\
  $I_3^F|\bar{p}>=-\frac{1}{2}|\bar{p}>, I_3^F|\bar{n}>=\frac{1}{2}
|\bar{n}>$ for B=-1.

The B=0 space is 6 dimensional, with elements like $|0>,
a^{\dagger}_1b^{\dagger}_2|0>...$.\\
In  $A^{\dagger}_{\mu}A_{\nu}$ notation, $|0>\rightarrow|3,4>,
a_1^{\dagger}b_2^{\dagger}|0>\rightarrow|3,1>\\
a_1^{\dagger}b_1^{\dagger}a_2^{\dagger}b_2^{\dagger}|0>
\rightarrow|1,2>...$.

There are 3 isospin 0 states:
\begin{eqnarray*}
|I=0>_1&=&|3,4>(=|0>)\\
|I=0>_2&=&|1,2>(=a_1^{\dagger}b_1^{\dagger}a_2^{\dagger}b_2^{\dagger}|0>)\\
|I=0>_3&=&\frac{1}{\sqrt{2}}(|1,4>-||2,3>)\\
&&(=\frac{1}{\sqrt{2}}(a_1^{\dagger}b_1^{\dagger}|0>-a_2^{\dagger}
b_2^{\dagger}|0>)
\end{eqnarray*}\\

Given the boson and fermion Hilbert spaces, we wish to construct a
fermion-boson trilinear coupling $H_I$,, which is an isospin invariant:
\begin{eqnarray}
I_3&=&I_3^F+I_3^B=\mathcal{A}\otimes I+I\otimes (J_{11}-J_{-1,-1})\\
I_+&=&I_+^F+I_+^B=\mathcal{A}\otimes
I+I\otimes\sqrt{2}(J_{1,0}+J_{0,-1})\\
H_0&=&\mathcal{A}(E)\otimes I+I\otimes\sum J_{k,k},
 E=\left[\begin{array}{cc}I&0\\0&-I\\
\end{array}\right]\\
H_I&=&\alpha\sum(\mathcal{A}(X_k)\otimes
J^-_k+\mathcal{A}(X^{\dagger}_k)\otimes J^+_k),
 \end{eqnarray}
with the matrices $X_k,X_k^{\dagger}$ determined by $[I_3,H]=[I_+,H]=0$;

this gives 

\begin{eqnarray*}
X_1&=&-X_{-1}^{\dagger}=\left[\begin{array}{cc}
\tau_+&\tau_+\\\tau_+&\tau_+\\
\end{array}\right]\\
X_0&=&X_0^{\dagger}=-\sqrt{2}\left[\begin{array}{cc}
\tau_3&\tau_3\\\tau_3&\tau_3\\
\end{array}\right]\\
X_{-1}&=&-X_1^{\dagger}=-\left[\begin{array}{cc}
\tau_-&\tau_-\\\tau_-&\tau_-\\
\end{array}\right]
\end{eqnarray*}
and the interacting Hamiltonian has the form
\begin{eqnarray}
H_I&=&\mathcal{A}(X_1)\otimes (J_1^- +J_{-1}^+)+\mathcal{A}(X_0)\otimes
(J_0^+-J_0^-)\nonumber\\
&&+\mathcal{A}(X_{-1})\otimes (J_{-1}^-+J_1^+)\\
&=&(A_1^{\dagger}+A_3^{\dagger})(A_2+A_4)\otimes(J_1^--J_{-1}^{\dagger})
\nonumber\\
&&-\frac{1}{\sqrt{2}}(A_1^{\dagger}+A_3^{\dagger})(A_1+A_3)\otimes(J_0^-
+J_{0}^{\dagger})\nonumber\\
&&+\frac{1}{\sqrt{2}}(A_2^{\dagger}+A_4^{\dagger})(A_2+A_4)\otimes(J_0^-+
J_{0}^{\dagger})\nonumber\\
&&+(A_2^{\dagger}+A_4^{\dagger})(A_1+A_3)\otimes(-J_{-1}^-+J_{1}^{\dagger}).
\end{eqnarray}

Given the isospin Hamiltonian, Eqs.(75),(76), we wish to solve the "vacuum"
problem
$H|\Psi>=\lambda_{min}|\Psi>$ for the "trilinear interaction" on the B=0, I=0
space.  M=1 generates 4 isospin 0 states;  in particular
\begin{eqnarray*}
|I=0>_4&=&\sum C^{0,1,1}_{0,m,-m} |I=1,m>_F\otimes |I=1,-m>_B\\
&=&\frac{1}{\sqrt{3}}[|1,3>\otimes
z_{-1}-\frac{1}{\sqrt{2}}(|1,4>+|2,3>)\otimes z_0\\
&&+|2,4>\otimes z_1];\\
H_I|I=0>_1&=&H_I(|3,4>\otimes w)\\
&=&\sqrt{3}|I=0>_4,\\
H_I|I=0>_2&=&H_I (|1,2>\otimes w)\\
&=&-\sqrt{3}|I=0>_4,\\
H_I|I=0>_3&=&H_I\frac{1}{\sqrt{2}}(|2,3>-|1,4>)\otimes w)\\&=&0,\\
H_I|I=0>_4&=&\sqrt{3}(|I=0>_1-|I=0>_2).
\end{eqnarray*}

Then the Hamiltonian matrix to be diagonalized is
\begin{eqnarray}
H&=&\left[\begin{array}{cccc}-2&0&0&\sqrt{3}\alpha\\0&2&0&-\sqrt{3}\alpha\\
0&0&0&0\\\sqrt{3}\alpha&-\sqrt{3}\alpha&0&0\\
\end{array}\right]
\end{eqnarray}
and has lowest eigenvalue $\lambda_{min}(\alpha)=-\sqrt{4+6\alpha^2.}$ 
For $\alpha \gg 1$ the eigenvalue goes as a negative constant times
$\alpha$, consistent with the behavior of trilinear couplings discussed
in Appendix B.  A factor 2 should be added to the free Hamiltonian, eq.(75);  then
the minimum eigenvalue is zero when the bare coupling is zero.

To model QCD behavior a quartic boson self-coupling term should be added. 
Though there is not a sufficiently rich isospin 0 quartic operator that
mimics the anharmonic oscillator, a possibility is
\begin{eqnarray*}
H_{self}&=&\alpha^2(\sum J^+_k J_k^-)^2;\\
\sum J_k^+ J_k^-|I=0>_4&=&3|I=0>_4,
\end{eqnarray*}
so the Hamiltonian to be diagonalized now is
\begin{eqnarray}
H&=&H_0+H_I+H_{self}\\
&=&\left[\begin{array}{cccc}-2&0&0&\sqrt{3}\alpha\\0&2&0&-\sqrt{3}\alpha\\
0&0&0&0\\\sqrt{3}\alpha&-\sqrt{3}\alpha&0&3\alpha^2\\
\end{array}\right];
\end{eqnarray}
in this simple model the "self-coupling" is not sufficient to produce an eigenvalue zero for the lowest
eigenvalue.  In the next paper
models with many velocity modes will be analyzed, adjoined with an internal
symmetry of the type analyzed in this section, to see if "self-coupling" is able to 
produce a zero eigenvalue for a particular value of the bare coupling constant.

\section{Conclusion}

This paper is the first in a series of papers that explore the possibility of approximating solutions of
quantum field theories through the use of inductive \cite{dd} and contractive limits \cite{i}.  The starting
point is an algebra of operators bilinear in fermion and antifermion creation and
annihilation operators which generate a unitary Lie algebra.  For infinite degree of
freedom systems there is a very large class of representations of this unitary algebra.
The idea is to explore the inductive limit of nested finite dimensional unitary subalgebras of
the full infinite dimensional algebra.

But before this inductive limit can be explored it is first necessary to deal with the boson 
algebra.  Here also there is an algebra of operators which, for infinite degree of freedom
systems, have a rich representation structure that can be explored as an inductive
limit.  The problem that aries for bosons is that, even before considering an inductive
limit, it is necessary to deal with the fact that in any mode, there can be indefinitely large
numbers of bosons.  One of the main goals of this paper has been to show how
contraction limits provide a natural way of approximating indefinitely large numbers
of bosons.
That is, if the bosonic Lie algebra, the semi-direct product of the Heisenberg
  group with a unitary group (see Eqs.(25) through (29)), is replaced by the
  compact group U(N+1), all of the irreps, and in particular the bosonic irreps
  of the form (M, 0,...,0)\cite{g}, are finite dimensional.  Here M is an irrep label that
  specifies the maximum number of allowed bosons in all of the modes;  as
  seen in Eq.(55), the raising operator, $J^+_i$ annihilates the state with
  the maximum number of bosons.  This is to be contrasted with simply
  putting a cutoff on the maximum number of allowed bosons, for here it
  is the Lie algebra itself that keeps the number of bosons finite.  When the
  contraction parameter $\rho$ goes to zero as the irrep label M goes to 
  infinity, such that M$\rho^2$=1, as shown in section 4, the full boson
  algebra is recovered.  Appendix A shows how the contractive limit is obtained for
   a simple exactly solvable model.  Further, a simple one-mode
  model based on the quantum anharmonic oscillator shows that full
  convergence to the limit for the ground state already occurs for M about 100;
  more work is required to see how large M must be for systems of many modes.
  
  The philosophy underlying these papers is to preserve as many symmetries as possible, even 
when the underlying variables are discrete and range over finite values.  Symmetries
 here include not only the usual Lorentz and internal symmetries, but also the
 unitary symmetries leading to inductive limits generated by fermion-antifermion bilinears (in creation and annihilation operators) and boson symmetries, with both inductive and contractive limits;  then interactions
  are given by products of generators of these Lie algebras, which
  in turn generate the four-momentum operator as the fundamental operator.
  
  The context of this paper is point form quantum field theory\cite{c}, and the basic equations
  to be satisfied are the point form equations (1) and (2).  In the point form all interactions
  are in the four-momentum operator and the Lorentz generators are kinematic.  Eigenvalues
  of the four-momentum operator, Eq. (3) generate the observables, the physical vacuum, the mass spectrum, and scattering
  states.  Lorentz transformations are automorphisms on the algebra of fermion and boson operators;  when the 
  algebra is finite dimensional (for a finite number of velocity modes) the Lorentz
  transformations are realized as a finite subgroup of the permutation group.
  
  A second goal of this paper has been to study generic properties of
  Hamiltonians with trilinear coupling, such as Eq.(16).  The motivation
  here is to study the vacuum properties of such Hamiltonians, in preparation
  for a study of the vacuum structure of the full four-momentum operator.
  A vacuum solution to the eigenvalue equation (3) means that in a suitable 
  representation space of the algebra of operators, the vector $|\Omega>$ must satisfy
  \begin{eqnarray}
  P^{\mu}|\Omega>&=&0,\\
  U_{\Lambda}|\Omega>&=&|\Omega>;
  \end{eqnarray}
  what is of interest here is that if the point form equations (1) and (2)
  are satisfied, it is not possible to add  constants to the four-momentum
  operator and still satisfy Lorentz covariance, Eq.(2).  Now the free four-momentum
  operator satisfies the above vacuum equations, where the vacuum
  is the usual Fock vacuum.  As shown in Appendix B, the generic structure of Hamiltonians
  with trilinear couplings is that the vacuum energy decreases linearly with
  the bare coupling constant.
  This means there can be no solution to the vacuum problem, Eqs.(82),(83), with just
  trilinear couplings.  Other interactions such as a boson self-interaction
  are necessary for the vacuum solution to give zero when acted on by
  the four-momentum operator.  Section 5 presented a simple one-mode
  model in which the boson self-interaction is given by the quantum anharmonic
  oscillator.  When combined with a trilinear coupling, the eigenvalue
  problem for small values of M gives nontrivial solutions to the vacuum problem.
  This raises the possibility of fine tuning the bare coupling constant, though, of course,
  much work remains to see whether this sort of fine tuning persists in the
  inductive limit.
  
  The longer term goal of this series of papers is to investigate the sorts
  of problems raised above when the number of velocity modes gets large, and in particular 
  to investigate the nature of the physical
  vacuum that is generated by the QCD Hamiltonian.  Here also there will be an
  interplay between trilinear interactions, of the quarks with the gluons, and
  gluon self-interaction terms.
  
  Acknowledgement:  I wish to thank Fritz Coester for many stimulating conversations.
I also thank Addison Stark and Kevin Murphy for results and discussions on the anharmonic
oscillator and one-mode trilinear coupling Hamiltonians.

\section{Appendix A:  Exactly Solvable Bosonic Contraction Model}
Consider the exactly solvable boson Hamiltonian, 
\begin{eqnarray}
H&=&\sum e_ic^{\dagger}_ic_i+\alpha\sum (
D_ic_i+D_i^{\ast}c_i^{\dagger})\\
&=&\sum e_i(c_i^{\dagger}+\frac{\alpha D_i}{e_i})(c_i+\frac{\alpha
D_i^{\ast}}{e_i})-\alpha^2\sum\frac{|D_i|^2}{e_i};\\
\lambda(gnd)&=&-\alpha^2\sum\frac{|D_i|^2}{e_i}.
\end{eqnarray}
In Eq.(84) the fermion operators in Eq.(18) are replaced by constants $D_i$.  As
is easily checked, $c_i+\frac{\alpha D_i^{\ast}}{e_i}$ also satisfies
bosonic commutation relations with its adjoint, and hence, the spectrum
of $H$ is $\sum e_i n_i -\alpha^2\frac{|D_i|^2}{e_i}$;  the ground state
is given by all the n's zero, which is Eq.(86).

The Hamiltonian $H_M$ which contracts to the Hamiltonian {H} of Eq.(84)
is made out of U(N+1) Lie algebra elements; the goal of this appendix
is to show that the ground state of $H_M$ contracts to the ground state
of $H$:

\begin{eqnarray}
H_M&=&\sum e_i\tilde{J}_{ii}+\alpha\sum
(D_i\tilde{J}_i^-+D_i^{\ast}\tilde{J}_i^+)\\ &=&Tr\mathcal{M}\mathcal{J}\\
&=&TrU^{\dagger}\Lambda U\mathcal{J}\\
&=&Tr\Lambda U\mathcal{J}U^{\dagger}\\
&=&\sum\Lambda_iJ_{ii}^{'}+\Lambda_{N+1}J_2^{'};
\end{eqnarray}

here the matrix
$\mathcal{M}=\left(\begin{array}{cc}e_i\delta_{ij}&\alpha
D_i\\\alpha D_j^{\ast}&0\end{array}
\right)$
and
$\mathcal{J}=\left(\begin{array}{cc}\tilde{J}_{ij}&\tilde{J}_i^+\\
\tilde{J}_j^-&\tilde{J}_2
\end{array}\right)=U^{\dagger}\mathcal{J}^{'}U$
is an automorphism of the U(N+1) Lie algebra.

Corresponding to the change of the Lie algebra basis is a change of the
representation basis:
\begin{eqnarray}
<z,w|M\vec{n}>^{'}&=& <(z,w)U|M\vec{n}>;\\
H_M|M\vec{n}>^{'}&=&(\sum\Lambda_in_i+\Lambda_{N+1}(M-\sum
n_i))|M\vec{n}>^{'}\\
&=&(\sum(\Lambda_i-\Lambda_{N+1})n_i+\Lambda_{N+1}M)|M\vec{n}>^{'};\\
H_M|M\vec{0}>^{'}&=&M\Lambda_{N+1}|M\vec{0}>^{'}
\end{eqnarray}
is the ground state ( with eigenvalues ordered by
$\Lambda_1>...>\Lambda_N>\Lambda_{N+1}$).

It is simplest to compute the contraction limit when there is only one
mode (N=1); then
$\mathcal{M}=\left(\begin{array}{cc}e&\alpha
D\\\alpha D^{\ast}&0\end{array}\right)$ with eigenvalues
$\Lambda_{\pm}=\frac{e\pm\sqrt{e^2+4\alpha^2|D|^2}}{2}$. 
The contraction limit is obtained by replacing $\alpha$ with
$\alpha\rho$ and letting $\rho\rightarrow0, M\rightarrow\infty$
such that $M\rho^2=1$.  Inserting the contraction parameter into the
miniumum eigenvalue,  Eq.(95), gives $M\Lambda_-=\frac{eM}{2}
(1-\sqrt{1+\frac{4\alpha^2|D|^2}{e^2}\rho^2})\rightarrow
-\frac{\alpha^2|D|^2}{e}$ as the ground state eigenvalue in the infinite
boson limit.  This result agrees with the ground state, Eq.(86) for one
mode.

\section{Appendix B: The Ground State Eigenvalue in the Strong Coupling Limit}
The goal of this appendix is to show that the Hamiltonian for trilinear fermion-boson
interactions has the property that the ground state goes as a linear function of the
bare coupling constant $\alpha$, for $\alpha \gg 1$. 
Moreover, the constant multiplying $\alpha$ is negative. The general form
of the Hamiltonian is given in Eq.(18); here it is assumed that the matrices
$X^0_i$ commute for different modes, $[X_i^0, X_j^0]=0$:
\begin{eqnarray}
H&=&\sum
e_i(a^{\dagger}_ia_i+b^{\dagger}_ib_i+\kappa c^{\dagger}_ic_i)
+\alpha\sum\mathcal{A}(X_i^0)c_i+
\mathcal{A}(X_i^0)^{\dagger}c_i^{\dagger};\\
\frac{H}{\alpha}&=&\sum c^{\dagger}_ic_i
+\sum\mathcal{A}(X_i^0)c_i+
\mathcal{A}(X_i^0)^{\dagger}c_i^{\dagger}-\sum c^{\dagger}_ic_i\nonumber\\
&&+\frac{1}{\alpha}\sum
e_i(a^{\dagger}_ia_i+b^{\dagger}_ib_i+\kappa c^{\dagger}_ic_i)\\
 &=&\sum
(c^{\dagger}_i+\mathcal{A}(X_i^0))(c_i+\mathcal{A}(X_i^0)^{\dagger})
-\sum\mathcal{A}(X_i^0)\mathcal{A}(X_i^0)^{\dagger}\\
&&-\sum c^{\dagger}_ic_i+\frac{1}{\alpha}\sum
e_i(a^{\dagger}_ia_i+b^{\dagger}_ib_i+\kappa c^{\dagger}_ic_i).
\end{eqnarray}
The part of the Hamiltonian given in Eq.(98) has a ground state
eigenvector and eigenvalue given by
\begin{eqnarray}
|\Psi_0>&=&N
e^{-\sum\mathcal{A}(X_i^0)^{\dagger}c^{\dagger}_i}|\Lambda>\\
\mathcal{A}(X^0_k)^{\dagger}|\Psi_0>&=&-\Lambda_k^{\ast}|\Psi_0>,
\end{eqnarray}
where $N$ is a normalization factor such that $<\Psi_0|\Psi_0>=1$.
Further, $c_k|\Psi_0>=-\Lambda_k^{\ast}|\Psi_0>$, so that
$<\Psi_0|\mathcal{A}(X^0_k)\mathcal{A}(X^0_k)^{\dagger}+
c_k^{\dagger}c_k|\Psi_0>=2|\Lambda_k|^2$.  Finally, the part of the Hamiltonian
multiplying $\frac{1}{\alpha}$ in Eq.(97) is a nonnegative operator,
$< \Psi_0|\sum 
e_i(a^{\dagger}_ia_i+b^{\dagger}_ib_i+\kappa c^{\dagger}_ic_i)|\Psi_0>:=<\Psi_0|H_0|\Psi_0>\ge 0$,
so that, in first order perturbation theory, the lowest eigenvalue of $\frac{H}{\alpha}$ is
$-2\sum |\Lambda|^2+\frac{<\Psi_0|H_0|\Psi_0>}{\alpha}$;  it then follows that the lowest
eigenvalue of $H$ in first order perturbation theory is 
$-2\alpha \sum |\Lambda|^2+<\Psi_0|H_0|\Psi_0>$.  That is, as a function of the bare
coupling constant $\alpha$, it goes linearly with $\alpha$, with a negative slope.

\end{document}